\documentclass[prb,twocolumn,amsmath,superscriptaddress]{revtex4}

\usepackage{graphicx}

\begin{document}

\title{Spin current injection by intersubband transitions in quantum wells. }
\author{E. Ya. Sherman, Ali Najmaie, and J.E. Sipe\\
Department of Physics, University of Toronto,\\
60 St. George Street, Toronto, ON M5S 1A7, Canada}

\begin{abstract}
We show that a pure spin current can be injected in quantum wells by 
the absorption of linearly polarized infrared radiation, leading to transitions
between subbands. The magnitude and the direction 
of the spin current depend on the Dresselhaus and Rashba spin-orbit 
coupling constants and light frequency and, therefore, can be 
manipulated by changing the light frequency and/or applying an external bias
across the quantum well. The injected spin current should be observable
either as a voltage generated via the anomalous spin-Hall effect, or by spatially resolved pump-probe
optical spectroscopy.
\end{abstract}

\maketitle 

\widetext

Spin current is an interesting physical phenomenon in its own right, and 
could have application in the
delivery and transfer of electron spins in spintronics devices. From a
fundamental point of view, various issues raised in the theory of this
effect are far from being satisfactorily settled. As was shown by
Rashba \cite{Rashba03}, a spin current exists even in the equilibrium state
of a two-dimensional (2D) electron gas with spin-orbit (SO) coupling. The application
of an external electric field has been suggested as a strategy for driving the
system out of equilibrium and inducing a spin current exhibiting transport
effects. Mal'shukov \textit{et al.} \cite{Malshukov03} and 
Governale \textit{et al.}  \cite{Governale03} suggested
applying a time-dependent bias across a semiconducting heterostructure, 
thus modulating the strength of the SO coupling and  generating a spin
current. Murakami \textit{et al}. \cite{Murakami03} and 
Sinova \textit{et al}. \cite{Sinova04} have shown that an in-plane electric 
field can cause a spin current, leading to the "intrinsic spin-Hall effect". 
Another possibility
for the injection of spin current is coherently controlled optical
excitations between the valence and the conduction band, as predicted by
Bhat and Sipe \cite{Bhat00,Bhat04} and observed experimentally in bulk crystals \cite
{Stevens02,Hubner03} and quantum wells (QWs) \cite{Stevens03}. 

Here we show that a spin
current can be injected in QWs by infrared (IR) light absorption, driving 
transitions between different subbands. The injection of {\it spin-polarized electric current} 
in QWs due to intersubband
transitions caused by circularly polarized radiation 
has already been observed by Ganichev \textit{et al.} \cite{Ganichev03}. 
In contrast, here we investigate a {\it pure spin current}, where electrons 
moving in opposite directions have opposite orientations of spins, not accompanied by a net
electrical current. We show that the
strength and direction of this pure spin current can be manipulated by
modulating the SO coupling strength via applied bias \cite{Nitta97}
and/or adjusting the light frequency.

As an example we consider the (011) GaAs QW,
where the electron spins have a
considerable out-of plane component, thus making possible the observation of the pure
spin current by detecting the voltage generated via the anomalous spin-Hall effect \cite{Abakumov72,Bakun85}.
The first two subbands in the well are typically separated by the energy 
$\hbar\omega_{0}\approx 100$ meV; the exact value depends on the width of the QW, dopant
concentration, and the boundary conditions. The SO Hamiltonian for the (011) QW,
${H}_{\mathrm{SO}}={H}_{D}+{H}_{R},$ is the sum of a Dresselhaus 
term \cite{Dyakonov86}, ${H}_{D},$ originating from the unit cell
inversion asymmetry, and a Rashba term \cite{Rashba84}, ${H}_{R},$
originating from the asymmetric doping and/or a bias applied across the well:

\begin{equation}
{H}_{D}^{[n]}=\alpha _{D}^{[n]}k_{y}{\sigma}^{_{z}}F_{[n]}(\mathbf{k)%
},\qquad {H}_{R}^{[n]}=\alpha _{R}^{[n]}\left( {\sigma}^{x}k_{y}-%
{\sigma}^{y}k_{x}\right) ,
\end{equation}
where $n$ is the subband index, $\mathbf{k}=(k_{x},k_{y})$ is the in-plane wavevector of 
the electron envelope function, 
$F_{[n]}(\mathbf{k)=}1-\left( k_{y}^{2}-2k_{x}^{2}\right)\lambda_{[n]}^{2}$, where 
$\lambda_{[n]}$ depends on the QW width $w$, and the ${\sigma}^{i}$ are the
Pauli matrices. The $z-$axis is 
perpendicular to the QW plane and the in-plane axes are: $x=[100]$ and
$y=[0\overline{1}1]$. The parameters $\alpha _{D}^{[n]}$ and 
$\alpha _{R}^{[n]}$ depend on $n$; in the model of
rigid QW walls one has 
$\alpha_{D}^{[n]}=-\alpha_{0}n^{2}\left(\pi/w\right)^{2}/2$, where $\alpha _{0}$ is the
Dresselhaus constant for the bulk, and $\lambda_{[n]}=w/n\pi$ \cite{Dyakonov86}. 
The deviation of $F_{[n]}(\mathbf{k)}$ from unity becomes important at electron concentrations 
$N_{\mathrm{el}}\approx 10^{12}$ cm$^{-2}$.

The spin-related energy is given by 
$E_{\mathrm{SO}}^{[n]}\left( \mathbf{k}\right) 
=\sqrt{\left(\alpha _{D}^{[n]}k_{y}F_{[n]}(\mathbf{k)}\right)^{2}+
\left(\alpha_{R}^{[n]}k\right)^{2}},$ with ''up'' $(u)$
and ''down'' $(d)$ states having energies 
$E_{u,d}^{[n]}\left( \mathbf{k}\right)=\pm E_{\mathrm{SO}}^{[n]}\left(\mathbf{k}\right)$, 
and leads to the subband spectra: 
\begin{equation}
\varepsilon _{s_{1}}=
\frac{\hbar^{2}k^{2}}{2m}
\pm E_{\mathrm{SO}}^{[1]}\left( \mathbf{k}\right),
\qquad 
\varepsilon_{s_{2}}=\hbar \omega_{0}+
\frac{\hbar^{2}k^{2}}{2m}
\pm E_{\mathrm{SO}}^{[2]}\left( \mathbf{k}\right). 
\end{equation}
where $m$ is the electron effective mass and the indices $s_{1},s_{2}$
describe the $u(+)$ and $d(-)$ spin states in the subbands $n=1$ and $n=2$,
respectively. The corresponding spin eigenstates $\phi^{s_{n}}_{\bf k}$
result in expectation values of the spin components:

\begin{equation}
\langle\left.\phi^{s_{n}}_{\bf k}\right|\sigma^{z}\left|\phi^{s_{n}}_{\bf k}\right.\rangle=
\pm\frac{\alpha_{D}^{[n]}k_{y}F_{[n]}(\mathbf{k)}}{E_{\mathrm{SO}}^{[n]}\left(\mathbf{k}\right)},
\qquad 
\langle\left.\phi^{s_{n}}_{\bf k}\right|{\sigma}_{\|}\left|\phi^{s_{n}}_{\bf k}\right.\rangle=
\pm\frac{\alpha_{R}^{[n]}}{E_{\mathrm{SO}}^{[n]}\left(\mathbf{k}\right)}(k_{y},-k_{x}),
\end{equation}

where upper(lower) sign corresponds to the $u(d)$ state and $\mathbf{\sigma}_{\|}=(\sigma^{x},\sigma^{y})$.

There is not yet consensus in the literature on the fundamental description of spin current, 
and the effect of disorder on it, as discussed e.g., in Ref.\cite{SHE}; spin current is not a "true" current, in that its density 
does not satisfy a continuity equation describing the evolution of a
spin density \cite{Rashba03}. Nonetheless, we introduce a "physical" definition of spin current 
per electron as:
\begin{equation}
j_{\mu}^{\beta }\left(\mathbf{k,}s_{n}\right) =\frac{\hbar}{4}\cdot 
\langle\left.\phi^{s_{n}}_{\bf k}\right|v_{\mu }{\sigma}^{\beta }+
{\sigma}^{\beta }v_{\mu}\left|\phi^{s_{n}}_{\bf k}\right.\rangle ,
\end{equation}
where $\mu$ and $\beta $ are Cartesian indices.
Velocity components 
${v}_{i}=\partial{H}/\hbar\partial k_{i}$ are the sums of  normal 
${v}_{i,{\rm n}}=\hbar k_{i}/m$ and anomalous terms given in our 
model (Eq.(1)) by:

\begin{equation}
v_{x,{\rm an}}^{[n]}=
-\frac{\alpha_{R}^{[n]}}{\hbar}{\sigma}^{y}+
4k_{y}k_{x}\frac{\alpha_{D}^{[n]}}{\hbar }{\sigma}^{z}\lambda_{[n]}^{2},
\qquad 
v_{y,{\rm an}}^{[n]}=
\frac{\alpha _{D}^{[n]}}{\hbar }\left(1-\left(3k_{y}^{2}-2k_{x}^{2}\right)\lambda_{[n]}^{2}\right)
{\sigma}^{z}.
\end{equation}

Below we consider only the spin current components associated with the $z-$%
axis spin projection. First we calculate the equilibrium spin current at
typical experimental conditions, where only the first subband is occupied,  
and then find the changes induced by the
intersubband excitations. For this purpose we introduce the equilibrium
Fermi distribution function for two spin projections in the first subband: 
\begin{equation}
f_{\pm}(\mathbf{k})=
\frac{1}{\exp\left[\left(\hbar^{2}k^{2}/2m\pm E^{[1]}_{\mathrm{SO}}(\mathbf{k})-\mu\right)/k_{B}T\right]+1},
\end{equation}
where $\mu $ is the chemical potential for a given $N_{\mathrm{el}}$, and $k_{B}$ 
is the Boltzmann's constant. The spin current density component 
$J_{y}^{z}$ is the sum of the normal, $J_{y,{\rm n}}^{z}$, and the anomalous,
$J_{y,{\rm an}}^{z}$, parts. By
integrating $j_{\mu }^{\beta}\left(\mathbf{k,}s_{1}\right)$ over the equilibrium state we obtain: 

\begin{equation}
J_{y}^{z} = \frac{\hbar}{2}
\left\{
\frac{\alpha
_{D}^{[1]}}{\hbar }\int 
\left(f_{+}\left(\mathbf{k}\right) + f_{-}\left(\mathbf{k}\right)\right)
\left( 1-\left( 3k_{y}^{2}-2k_{x}^{2}\right)\lambda_{1}^{2}\right) 
\frac{d^{2}k}{\left(2\pi\right)^{2}} 
+\frac{\hbar}{m}
\int\left(f_{+}\left(\mathbf{k}\right)-f_{-}\left(\mathbf{k}\right)\right)
k_{y}\left.\langle\phi^{u_{1}}_{\bf k}\right|\sigma^{z}\left|\phi^{u_{1}}_{\bf k}\right.\rangle
\frac{d^{2}k}{\left(2\pi\right)^{2}}
\right\} ,
\end{equation}
where $\left.\langle\phi^{u_{1}}_{\bf k}\right|\sigma^{z}\left|\phi^{u_{1}}_{\bf k}\right.\rangle$ 
is defined in Eq.(3), and $J_{x}^{z}=0$ by symmetry. 
The contributions   $J_{y,{\rm an}}^{z}$ and $J_{y,{\rm n}}^{z}$ (first and second term in Eq.(7), respectively) 
almost cancel each other. 
At $T=0$ each of them is close in absolute value to 
$\left|\alpha_{D}^{[1]}\right|N_{\mathrm{el}}/2\left(1-\pi N_{\mathrm{el}}\lambda_{[n]}^{2}/2\right)$,
and 
$J_{y}^{z}/J_{y,\mathrm{an}}^{z}\approx\left(m\alpha_{R}^{[1]}/\hbar^{2}k_{F}\right)^{2}\ll 1$
where $\hbar k_{F}$ is the Fermi momentum (see also Rashba \cite{Rashba03}).
We show $J_{y}^{z}/J_{y,\mathrm{an}}^{z}$ as a function of $N_{\rm el}$ in Fig. 1. 

Now we can investigate the spin current injection by linearly-polarized IR radiation due
to the intersubband transitions, as shown in Fig. 2a.
The external field is a pulse 
$\mathbf{E}(t)=\mathcal{E}(t)\exp\left(-i\omega t\right)+c.c.$ 
with the carrier frequency $\omega $, and
slowly varying amplitude $\mathcal{E}(t)$ of duration $\tau$.
We consider oblique incidence with $\mathcal{E}(t)$ lying in the plane
of incidence. The radiation frequency $\omega $ is close to $\omega _{0}$,
with a detuning $\Omega=\omega-\omega_{0}$, 
such that it can cause transitions between the subbands, with 
$\hbar\Omega$ being of the order of few meV.
For $\tau \gg \omega ^{-1}$ the exact shape of the pulse has no influence on our results; however, to
have the possibility of momentum-selective excitations as  shown in Fig.2a one
needs sufficiently long pulses, with $\tau >\hbar /\alpha _{D}k_{F}\approx 1$
ps, for $\alpha _{D}$ $\approx 10^{-9}$ eV$\cdot $cm, a typical value of the
Dresselhaus coupling \cite{Dyakonov86}. This condition also implies applicability of Fermi's
Golden Rule, since the pulse contains many periods of the field oscillations.
Since $\alpha_{D}^{[n]},$ $\alpha_{R}^{[n]}$ and, in turn, the
spin states and anomalous velocities depend on the subband, the intersubband
transitions can cause the injection of a spin current. The ratio 
$\alpha_{D}^{[n]}F_{[n]}(\mathbf{k})/\alpha_{R}^{[n]}$,  
which determines the direction of the effective SO field acting on the spin, depends on the subband. 
Therefore, the spin states in different subbands are not mutually orthogonal, so 
$\left.\langle\phi^{s_{1}}_{\bf k}\right|\phi^{s_{2}}_{\bf k}\rangle\neq 0$, and, 
''spin-flip'' transitions $u\longleftrightarrow d$ are allowed with linearly
polarized IR light absorption. 
The transitions $\phi^{s_{1}}_{\bf k}\rightarrow\phi^{s_{2}}_{\bf k}$
occur in the vicinity of the resonance curves in the momentum space, 
determined by the $\mathbf{k}=\mathbf{k}_{r}^{s_2,s_1}(\Omega)$ where 
$\mathbf{k}_{r}^{s_2,s_1}(\Omega)$ is specified by the constraint 
of energy conservation. For a given $\Omega$ there are in fact two such curves. 
In our case $E_{\mathrm{SO}}^{[2]}\left(\mathbf{k}\right)>E_{\mathrm{SO}}^{[1]}\left(\mathbf{k}\right)$ 
for all $k$. Therefore, for $\Omega>0$
the transitions $d\rightarrow u$ and $u\rightarrow u$ are allowed, while for 
$\Omega <0$ we obtain $d\rightarrow d$ and $u\rightarrow d$ transitions.
As one can see in Figs. 2a and 2b, $k_{r}^{s_2,s_1}(\Omega)$ 
is larger for the ''spin-conserving'' than for the ''spin-flip''-transitions.

The transition matrix elements depend on the spin states in both subbands,
and can be factorized in the dipole approximation as: 

\begin{equation}
M\left(\mathbf{k}_{r}^{s_2,s_1}(\Omega)\right) =
\mathcal{E}e\frac{\sin 2\theta_{0}}{\epsilon\cos\theta_{0}+\sqrt{\epsilon }\cos \theta _{1}}
\left\langle\varphi^{(2)}(z)\right| z\left|\varphi ^{(1)}(z)\right\rangle
\left\langle\phi^{s_{2}}_{\bf k}\right|\left.\phi^{s_{1}}_{\bf k}\right\rangle,
\qquad 
\end{equation}

where $\theta_{0}$ and $\theta_{1}$ are, respectively, the incidence and refraction angles,
$\sin\theta_{1}=\sin\theta_{0}/\sqrt{\epsilon}$, $\epsilon$ is the dielectric constant, $e$ is the electron charge, 
and $\varphi^{(1)}(z)$, $\varphi^{(2)}(z)$ are the envelope electron wavefunctions in the
subbands $n=1$ and $n=2,$ respectively. 
A transfer of one electron to the second subband injects a spin current:
\begin{equation}
\Delta j_{y}^{z}\left( \mathbf{k;}s_{2},s_{1}\right) =
j_{y}^{z}\left(\mathbf{k,}s_{2}\right)-j_{y}^{z}\left(\mathbf{k,}s_{1}\right),
\end{equation}
where we neglect the small photon momentum. The incident radiation injects the concentration 
of electrons in the second subband $N_{2}$, with a rate $dN_{2}\left(\Omega\right)/dt$ and,
correspondingly, drives the spin current density
component with the rate $d\Delta J_{y}^{z}\left(\Omega\right)/dt.$ 
The injection rates can be written as:

\begin{equation}
\frac{d\Delta J_{y}^{z}\left(\Omega\right)}{dt}=
\frac{\hbar }{2}\zeta (\Omega )
\frac{dN_{2}\left(\Omega\right)}{dt},\qquad 
\frac{dN_{2}\left(\Omega \right)}{dt}=
\frac{\xi(\Omega)}{\hbar\omega}\left\langle S\right\rangle,
\end{equation}
where $\zeta (\Omega )$ characterizes the effective speed of electrons forming the
pure spin current, $\left\langle S\right\rangle =(c/2\pi)\mathcal{E}^{2}$ is
the radiation power per unit area, and $\xi(\Omega)$ is a dimensionless function.

Within Fermi's Golden Rule the speed characterizing the spin injection is obtained
as: 

\begin{equation}
\frac{\hbar}{2}\zeta(\Omega)=
\frac{\sum\limits_{s_{1},s_{2}}\displaystyle{\int}
f_{s_{1}}\left(\mathbf{k}\right)\left|\langle\phi_{\bf k}^{s_{2}}\right|\left.
\phi_{\bf k}^{s_{1}}\rangle\right|^{2}\Delta j_{y}^{z}
\left(\mathbf{k;}s_{2},s_{1}\right)
dk_{r}^{s_2,s_1}(\Omega)/v^{s_{2},s_{1}}_{\mathbf{k}}}
{\sum\limits_{s_{1},s_{2}}
\displaystyle{\int}f_{s_{1}}\left(\mathbf{k}\right)
\left|\langle\phi_{\bf k}^{s_{2}}\right|\left.\phi_{\bf k}^{s_{1}}\rangle \right|^{2}
dk_{r}^{s_2,s_1}(\Omega)/v^{s_{2},s_{1}}_{\mathbf{k}}},
\end{equation}

with the velocity associated with the joint density of states given by:
\begin{equation}
\mathbf{v}^{s_{2},s_{1}}_{\mathbf{k}} =
\frac{\partial}{\hbar\partial\mathbf{k}}
\left(\varepsilon_{s_{2}}-\varepsilon_{s_{1}}\right).
\end{equation}
The integration in Eq.(11) is performed along the resonance curves.
With the increase of $\left|\Omega\right|$, 
$k_{r}^{s_2,s_1}(\Omega)$ increases and eventually arrives at regions of small electron occupancy,
as can be seen in Fig. 2b. Hence, $d\Delta J_{y}^{z}\left(\Omega\right)/dt$ and 
$dN_{2}/dt$ become small at $\hbar\left|\Omega\right|$ larger than some
critical $\hbar\Omega_{\rm c}$ (a few meV) determined by the condition 
$\min k_{r}^{s_2,s_1}(\Omega_{\rm c})>k_{0}$, 
where $k_{0}=k_{F}$ or  $k_{0}=\sqrt{mk_{B}T}/\hbar$ in the
degenerate and non-degenerate gas, respectively.

The photoinduced spin current is the sum of normal 
$\Delta J_{y,\mathrm{n}}^{z}$ and anomalous $\Delta J_{y,\mathrm{an}}^{z}$ contributions,
each containing spin-flip $\left(s_{1}\neq s_{2}\right) $ and spin
conserving $\left( s_{1}=s_{2}\right) $ terms. The anomalous spin-conserving
term is of the order of 
$\left(\alpha _{D}^{[2]}-\alpha _{D}^{[1]}\right)N_{2},$
while the other terms depend on the difference of the ratio 
$\alpha_{R}^{[2]}/\alpha_{D}^{[2]}-\alpha_{R}^{[1]}/\alpha _{D}^{[1]}$ 
and $\lambda_{[n]}$. An estimate of the relative contributions is: 
\begin{equation}
\frac{\Delta J_{y,\mathrm{n}}^{z}\left( s_{1}=s_{2}\right)}
{\Delta J_{y,\mathrm{an}}^{z}\left( s_{1}=s_{2}\right)}
\approx
\frac{\hbar ^{2}k_{F}}{m\alpha _{D}}
\left[
\left(\frac{\alpha _{R}^{[2]}}{\alpha_{D}^{[2]}}\right)^{2}
-\left(\frac{\alpha_{R}^{[1]}}{\alpha_{D}^{[1]}}\right)^{2}
\right] .
\end{equation}
Due to a large prefactor $\hbar^{2}k_{F}/m\alpha_{D},$ which is the ratio
of the normal and anomalous velocities, the normal term can be large 
and lead to a change in the sign of the spin current
at particular light frequencies, as seen in Figs.3a and 3b.
In Fig.3a we present the speed $\zeta(\Omega)$, while 
in Fig.3b we show the normal and anomalous parts of the injected spin current density. 
The spin-flip contribution in both the normal and anomalous terms is much smaller than the
''spin-conserving'' one. Recently, Golub 
\cite{Golub03} demonstrated that 
the direction of electric current induced by interband light absorption in QWs
can depend on the light frequency. In his scenario the change occurs as
new subbands are accessed, and thus appears on a scale of 100 meV. In our scenario
for pure spin current injection, the change occurs on a much smaller scale. 

Now we estimate the magnitude of the injected spin current assuming that the
contributions of the anomalous and normal terms are of the same order of
magnitude. Fig.4 presents the efficiency of the energy absorption 
$\xi(\Omega )$ (Eq.(10)). The concentration of the electrons excited to the second subband
can be estimated from Eqs.(8)-(10) as 
$N_{2}\approx 2\pi(e^{2}w^{2}k_{F}/\epsilon^{2}\hbar c\alpha_{D})\langle S\rangle\tau$.
At $\epsilon =12$, $k_{F}\approx 10^{6}$ cm$^{-1},w=100$ \AA, 
$\theta_{0}$ close to $\pi/4$ and $N_{\mathrm{el}}\approx 10^{12}$
cm$^{-2}$ we obtain: 
$N_{2}/N_{\mathrm{el}}\approx 10^{-6}$ $\left(\langle S\rangle/(\mathrm{W/cm}^{2})\right)\cdot\left(\tau/\mathrm{ps}\right)$. 
Under excitation of a 1\% fraction of electrons,
achieved at $\langle S\rangle\approx 10$ $\mathrm{kW/cm}^{2}$ and 
$\tau \approx 1$ $\mathrm{ps,}$ the corresponding effective current density 
$e\Delta J_{y}^{z}/\hbar\approx 1$ $\mathrm{mAmp/cm.}$ This is of the same magnitude 
as would be generated by the ac spin pumping in the $n=1$ subband, 
as proposed by Mal'shukov \textit{et al.} \cite{Malshukov03}, but the 
effect would operate on a nanosecond time scale, as opposed to the
picosecond time scale relevant here.

Having found the magnitude of the spin current, we discuss 
its experimental observation. A possible technique is the measurement
of the voltage generated by the anomalous spin-Hall effect due to
scattering of electrons by impurities. The spin current 
$\Delta J_{y}^{z}$ causes a spin-Hall bias $V_{sH}$ along the $x$ axis. Its
magnitude can be estimated as $V_{sH}\approx\tan(\theta_{sH})V_{\mathrm{eff}}$, 
where $\theta _{sH}$ is the spin-Hall angle and 
$V_{\mathrm{eff}}$ is the effective lateral bias that would cause a current 
density $e\Delta J_{y}^{z}/\hbar$. 
As follows from the discussion preceding Eq.(13), the 
corresponding current density is
of the order of  $eN_{2}(\alpha_{D}/\hbar)$. 
The bias $V_{\mathrm{eff}}$ that would cause this current density is: 
$V_{\mathrm{eff}}\approx L\left(N_{2}/N_{\mathrm{el}}\right)\alpha_{D}/\hbar\mu$,
where $\mu $ is the mobility, and $L\approx 1$ cm is the lateral size of the system. 
At $\mu \approx 10^{5}$ $\mathrm{cm}^{2}/(\mathrm{Vs})$, $N_{2}/N_{\mathrm{el}}\approx 10^{-3}$, and
$\alpha_{D}/\hbar\approx 10^{6}$ cm/s we obtain: 
$V_{\mathrm{eff}}/L\approx 10^{-2}$ $\mathrm{V/cm.}$ The spin-Hall angle was
estimated by Huang\textit{ et al.} \cite{Huang04} as $\theta_{sH}\approx 10^{-3}$, which would lead
to  $V_{sH}\approx 10^{-5}$ V. Their model assumed charged dopants embedded directly in the QW,
which considerably overestimates  the magnitude of the
effect when only a remote doping is present. For this reason,  $10^{-5}$ V is clearly
an upper estimate of the spin Hall bias.  Nonetheless, even a bias smaller by two orders of magnitude than
this would be experimentally accessible \cite{Bakun85}.

Another possibility for observing the pure spin current is spatially resolved
pump-probe spectroscopy, as applied by H\"{u}bner {\it et al.} \cite{Hubner03} and
Stevens {\it et al.} \cite{Stevens03} to
investigate the spin current injected by interband transitions. In those
experiments the centers of the spin-up and spin-down of excited electron
distribution were
separated by approximately 20 nm. In the experimental situation considered
here, the spin-polarized spots can be separated by distances of the order of the electron free path 
$\ell\approx\left(\hbar k_{F}/m\right)\tau_{k},$ with $\tau_{k}$ being the momentum
relaxation time. At mobility 
$\mu\approx 10^{5}$ $\mathrm{cm}^{2}/(\mathrm{Vs}),$ 
one obtains $\ell \approx 10^{3}$ nm, and so a possible approach would be
to observe this separation experimentally by using a linearly 
polarized IR light as a pump and circularly polarized light as a probe 
of the spin-dependent transmission.  In a real sample, of course, we have to expect 
some inhomogeneity in the spin-orbit interaction due to quantum well thickness variations, 
dopant fluctuations, inhomogeneous strain, and the like \cite{Sherman03}.  
We are currently investigating the consequences of such inhomogeneity, and 
will return to it in a later communication.

To conclude, we have shown that a pure spin current can be injected in QWs
by IR intersubband absorption, calculated its magnitude, and found that
it could  be measured experimentally. The dependence of the spin current on the
light frequency, and on the Rashba SO coupling parameter, opens the 
possibility of its manipulation applying an external bias and by changing the
light frequency. The spin current should be observable by anomalous spin-Hall 
effect measurements or by pump-probe optical spectroscopy.

E.Y.S is grateful to the Austrian Science Fund for financial support. A.N. acknowledges
support from an Ontario Graduate Scholarship. This work was supported in part by the National
Science and Engineering Research Council or Canada (NSERC) and the DARPA SpinS program. 
We thank P. Marsden, H. van Driel, and J. Sinova for useful discussions.

\newpage

\begin{figure}[htb]
\begin{center}
\includegraphics[width=7.0cm]{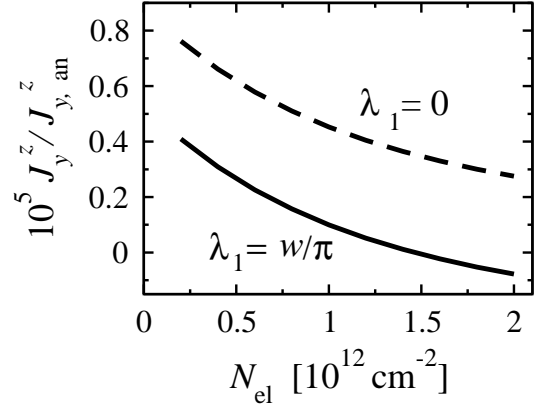}
\end{center}
\caption{ Spin current $J_{y}^{z}/J_{y,\mathrm{an}}^{z}$ as a
function of the electron concentration $N_{\mathrm{el}}$. Dashed curve : 
$\lambda_{1} =0$, solid curve $\lambda_{1} =w/\pi $, the QW width $w=80$ \AA. 
The parameters are: $\alpha _{D}^{[1]}$=-0.3$\times 10^{-9}$ eVcm, 
$\alpha_{R}^{[1]}$=-0.3$\alpha _{D}^{[1]}$, $k_{B}T$=25 meV,
$m=0.066m_{0},$ where $m_{0}$ is a free electron mass.
}
\end{figure}

\newpage 

\begin{figure}[]
\begin{center}
\includegraphics{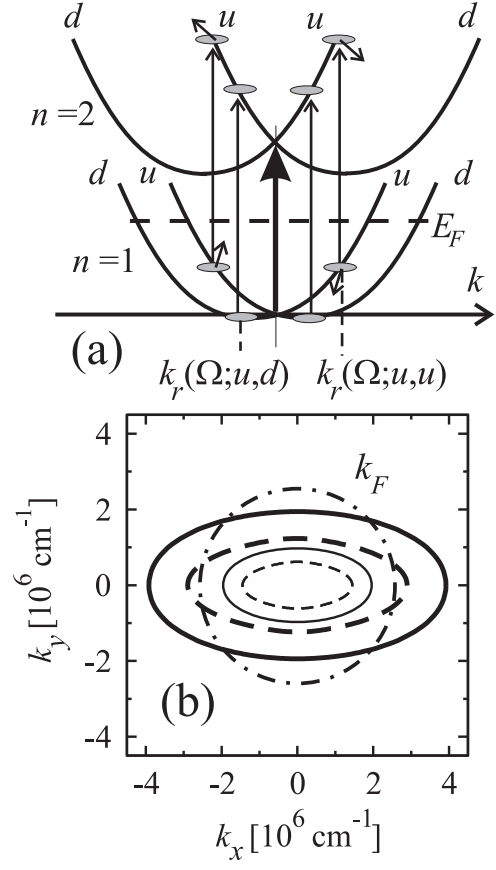}
\end{center}
\caption{ (a) The intersubband transitions leading to
the injection of pure spin current. Thick arrow line corresponds to 
$\Omega=0$. Thin arrow lines correspond to transitions at $\Omega>0$.
(b) Resonance curves $\mathbf{k}=\mathbf{k}_{r}^{s_2,s_1}(\Omega)$. 
Solid (dash) lines describe the spin-conserving (spin-flip) transitions. 
In each case the outer curve is for  $\hbar\Omega = 2$ meV and the inner curve for $\hbar\Omega = 1$ meV.
The circle marked as $k_F$ is the Fermi line at $N_{\rm el}=10^{12}$ cm$^{-2}$.  
The parameters are: 
$\alpha _{D}^{[1]}$=-0.3$\times 10^{-9}$ eVcm, 
$\alpha _{R}^{[1]}$=-0.3$\alpha _{D}^{[1]}$, 
$\alpha _{D}^{[2]}$=4$\alpha _{D}^{[1]}$, 
$\alpha_{R}^{[2]}$=-0.5$\alpha _{D}^{[2]}$, 
and $\lambda_{[n]}=w/n\pi$. 
}
\end{figure}

\newpage

\begin{figure}[htb]
\begin{center}
\includegraphics[width=7.0cm]{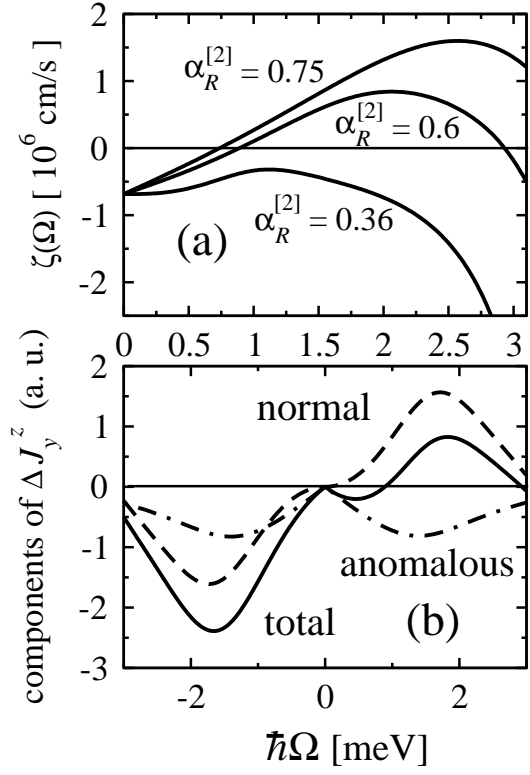}
\end{center}
\vspace{5cm}
\caption{(a) The speed $\zeta(\Omega)$
for different Rashba coupling constants $\alpha _{R}^{[2]}$ (values in 10$^{-9}$ eVcm 
units are presented near the curves). Other parameters are the same 
as in Fig.2, $N_{\mathrm{el}}=10^{12}$ cm$^{-2}$,
and $k_{B}T$=25 meV. The direction of the spin current can be altered by changing the Rashba parameter.
(b) Components of the induced spin current as the function of the photon
frequency for $\alpha _{R}^{[2]}=0.6\times 10^{-9}$ eVcm. 
}
\end{figure}

\newpage 

\begin{figure}[htb]
\begin{center}
\includegraphics[width=7.0cm]{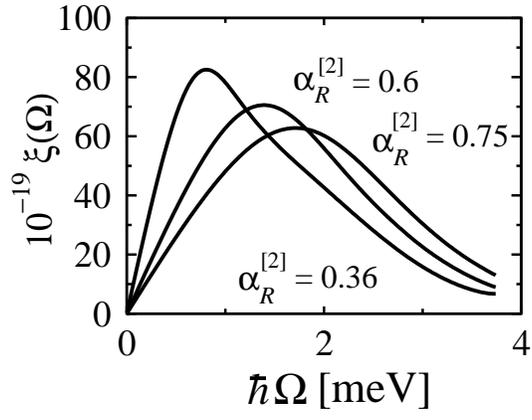}
\end{center}
\caption{$\Omega $-dependence of $\xi(\Omega)$ for different Rashba
parameters $\alpha _{R}^{[2]}$. The units of incident light power
density are $\mathrm{W/cm}^{2}$. We take an
incidence angle $\theta_{0}=\pi/4$,  $\epsilon=12$, and $w=80$ \AA.
The infinite barrier approximation is used for the calculation. 
}
\end{figure}

\end{document}